\documentstyle[amsmath,amsfonts]{article}
\def\be{\begin{equation}}
\def\ee{\end{equation}}
\def\ba{\begin{array}}
\def\ea{\end{array}}
\def\bea{\begin{eqnarray}}
\def\eea{\end{eqnarray}}
\def\d{{\rm d}}

\bibliography{plain}
\title{An Explicit Computation of the Bures Metric Over the Space of $N$-Dimensional Density
Matrices} \vspace{20mm}
\author{
  S. J. Akhtarshenas
\thanks{E-mail:akhtarshenas@phys.ui.ac.ir}
\\
{\small Department of Physics, University of Isfahan, Isfahan,
Iran } }
\begin{document}
\maketitle
%\vspace{15mm}
%\newpage

\begin{abstract}
The aim of this paper is to provide a method  for explicit
computation of the Bures metric over the space of $N$-level
quantum system, based on the coset parametrization of density
matrices.


\end{abstract}
%\pagebreak
%\vspace{7cm}
\section{Introduction}
Recent developments in the emerging field of quantum information
theory have received a great deal of attention in investigation of
the properties of the set of density matrices of an $N$-level
quantum system. In view of such considerable interest a lot of
work has been devoted to describe and parameterize density
matrices. Any density matrix of an $N$-level quantum system can be
expanded in terms of $N^2-1$ orthogonal generators of $SU(N)$
\cite{fano}, which is a generalization of the Bloch or coherence
vector representation for two-level systems. Boya et al
\cite{boya} have shown that the mixed state density matrices for
$N$-level quantum systems can be parameterized in terms of squared
components of an $(N-1)$-sphere and unitary matrices. By using the
Euler angle parametrization of $SU(3)$ group \cite{byrd1}, Byrd
and Slater \cite{byrd2} have presented a parametrization for
density matrices of three-level systems. An Euler angle-based
parametrization for density matrices of four-level systems is also
introduced in \cite{tilma1}, and has been used by Tilma and
Sudarshan \cite{tilma2} in order to study the entanglement
properties of the two-qubit system. A generalized Euler angle
parametrization for $SU(N)$ and $U(N)$ groups is given by Tilma
and Sudarshan \cite{tilma3,tilma4}. In a comprehensive analysis
\cite{karol}, \.{Z}yczkowski and S\l omczy\'nski analyzed the
geometrical properties of the set of mixed states for an arbitrary
$N$-level system and classified the space of density matrices.
Di\c{t}\v{a} \cite{dita1} has provided an explicit parametrization
for general $N$-dimensional Hermitian operators that may be
considered either as Hamiltonian or density matrices. The
parametrization is based on the factorization of $N\times N$
unitary matrices \cite{dita2}. A parametrization useful to study
the entanglement properties of two-qubit density matrices is also
introduced in \cite{akhtar1}, in which authors have shown that the
space of two-qubit density matrices can be characterized with
12-dimensional (as real manifold) space of complex orthogonal
group $SO(4,\mathbb{C})$ together with four positive Wootters's
numbers \cite{woot}, where, of course, the normalization condition
reduces the number of parameters to 15.

Efforts have been also made to study the geometry of density
matrices. In recent years, the Riemannian Bures metric
\cite{bures} has become an interesting subject for the
understanding of the geometry of quantum state space. It is the
quantum analog of Fisher information in classical statistics, i.e.
in the subspace of diagonal matrices it induces the statistical
distance \cite{braun}. The Bures measure is monotone in the sense
that it does not increase under the action of completely positive,
trace preserving maps \cite{petz}. It is, indeed, minimal among
all monotone metrics and its extension to pure state is exactly
the Fubini-Study metric \cite{petz}. The Bures distance between
any two mixed states $\rho_1$ and $\rho_2$ is a function of their
fidelity $F(\rho_1,\rho_2)$ \cite{uhlmann,jozsa}
\begin{equation}\label{fidelity}
\textmd{d}_{B}(\rho_1,\rho_2)=\sqrt{2-2\sqrt{F(\rho_1,\rho_2)}},\qquad
F(\rho_1,\rho_2)=
\left[\textmd{Tr}\left(\sqrt{\sqrt{\rho_1}\rho_2\sqrt{\rho_1}}\right)\right]^2.
\end{equation}
Fidelity allows one to characterize the closeness of the pair of
mixed states $\rho_1$ and $\rho_2$, so, it is an important concept
in quantum mechanics, quantum optics and quantum information
theory. An explicit formula for the infinitesimal Bures distance
between $\rho$ and $\rho+\textmd{d}\rho$ was found by H\"{u}bner
\cite{hubner}
\begin{equation}\label{hubner}
\textmd{d}_B(\rho,\rho+\textmd{d}\rho)^2=\frac{1}{2}\sum_{i,j=1}^{N}\frac{|\langle
\lambda_i|\textmd{d}\rho|\lambda_j\rangle|^2}{\lambda_i+\lambda_j},
\end{equation}
where $\lambda_j$ and $|\lambda_j\rangle$, $(j=1,2,\cdots,N)$
represent eigenvalues and eigenvectors of $\rho$, respectively.
Dittmann has derived several explicit formulas, that do not
require any diagonalization procedure, for Bures metric on the
manifold of finite-dimensional nonsingular density matrices
\cite{ditt1,ditt2}.

The probability measure induced by the Bures metric in the space
of mixed quantum states has been defined by Hall \cite{hall}. The
question of how many entangled or separable states are there in
the set of all quantum states is considered by \.{Z}yczkowski et
al in Refs. \cite{karol2,karol3}. Sommers and \.{Z}yczkowski
\cite{sommers} have computed the Bures volume of the
$(N^2-1)$-dimensional convex set and the $(N^2-2)$-dimensional
hyperarea of the density matrices of an $N$-level quantum system.
In a considerable work, Slater investigated the use of the volume
elements of the Bures metric as a natural measure over the
$(N^2-1)$-dimensional convex set of $N$-level density matrices to
determine or estimate the volume of separable states of the the
pairs of qubit-qubit \cite{slater2,slater3} and qubit-qutrit
\cite{slater4,slater5}.

By using the Dittmann formula \cite{ditt1} and the Euler-angle
parametrization \cite{byrd2}, Slater has computed the Bures metric
for the eight-dimensional state space of the three-level quantum
systems \cite{slater1}. In a similar work, but instead in terms of
the coset space  parametrization, we have very recently given an
explicit expression for the Bures metric of the space of
three-level quantum system \cite{akhtar2}. The coset
parametrization provides a geometrical description of the set of
density matrices and, as well as the Euler angle parametrization
does, eliminates any overparametrization of the density matrix
\cite{dita1}.

The aim of this paper is to provide a method for computation of
the Bures metric in $N$-level quantum systems, based on the
H\"{u}bner formula and the  coset parametrization of density
matrices. We also use the possibility of factorizing each coset
component  in terms of a diagonal phase matrix and an orthogonal
matrix \cite{dita2,chaturvedi}. The paper, therefore, can be
regarded as a further development in the explicit computation of
the Bures metric. We show that in the canonical coset
parametrization, the Bures metric matrix is divided into two
blocks, an $(N-1)$-dimensional diagonal matrix corresponding to
the eigenvalues coordinates, and an $N(N-1)$-dimensional matrix
corresponding to the coset coordinates.  It therefore provides a
factorization of the Bures measure on the space of density
matrices as the product of the measure on the space of eigenvalues
and the truncated Haar measure on the space of unitary matrices.
It is shown that the coset parametrization gives a compact
expression for all metric elements. The analytical expression for
Bures metric  can be used in computing the Bures volumes of
quantum states as well as to study the problem of what proportion
of the convex set of the bipartite systems is separable. The
results also enable the calculation of the minimum Bures distance
of a given density matrix from the convex set of separable states,
in order to quantify  entanglement of the state \cite{ved1,ved2}.

The paper is organized as follows: In section 2, the coset space
parametrization of an $N$-level density matrix is introduced.
Based on the parametrization, we provide in section 3 a formula
for computation of the Bures metric, explicitly.  The paper is
concluded in section 4 with a brief conclusion.

\section{Canonical coset parametrization of density matrices}
The state space of an $N-$level quantum system is identified with
the set of all $N\times N$ Hermitian positive semidefinite complex
matrices of trace unity, and comprises an $(N^2-1)$-dimensional
convex set ${\mathcal M}_N$. The total number of independent
variables needed to parameterize a density matrix $\rho$ is equal
to $N^2-1$, provided no degeneracy occurs. Let us denote the set
of all diagonal density matrices of an $N$-level quantum system by
${\mathcal D}_N$. An arbitrary element ${\rho^{(D)}\in {\mathcal
D}_N}$ can be written as
\begin{equation}\label{RhoD}
\rho^{(D)}=\textmd{diag}\{\lambda_1,\lambda_2,\cdots,\lambda_N\},\qquad
0\le\lambda_i\le 1, \qquad \sum_{i=1}^{N}\lambda_i=1,
\end{equation}
which simply denotes an $(N-1)$-dimensional simplex ${\mathcal
S}_{N-1}$ for the set of all diagonal density matrices.

A generic density matrix $\rho\in{\mathcal M}_N$ in an arbitrary
basis can be obtained as the orbit of points $\rho^{(D)}\in
{\mathcal D}_N$ under the action of the unitary group $U(N)$ as
\begin{equation}
\rho=U{\rho^{(D)}} U^{\dag}.
\end{equation}
Let $H$ be a maximum stability subgroup, i.e. a subgroup of $U(N)$
that consists of all the group elements $h$ that will leave the
diagonal state $\rho^{(D)}$ invariant,
\begin{equation}
h\rho^{(D)}h^\dag =\rho^{(D)}, \qquad h\in H, \qquad \rho^{(D)}\in
{\mathcal D}_N,
\end{equation}
that is, $H$ contains all elements of $U(N)$ that commute with
$\rho^{(D)}$. For every element $U\in U(N)$, there is a unique
decomposition of $U$ into a product of two group elements, one in
$H$ and the other in the quotient $G/H$ \cite{gilmore}, i.e.
\begin{equation}
U={\bf \Omega} \; h, \qquad U\in U(N), \qquad h\in H, \qquad {\bf
\Omega}\in U(N)/H.
\end{equation}
Therefore the action of an arbitrary group element $U\in U(N)$ on
the point $D\in {\mathcal D}_N$ is given by
\begin{equation}
\rho=U\rho^{(D)}U^\dag={\bf \Omega} h \rho^{(D)} h^\dag {\bf
\Omega}^\dag={\bf \Omega} \rho^{(D)} {\bf \Omega}^\dag.
\end{equation}
Since ${\mathcal D}_N$ consists of points with different degree of
degeneracy, the maximum stability subgroup will differ for
different $\rho^{(D)}\in {\mathcal D}_N$ \cite{karol}. Let $m_i$
denotes degree of degeneracy of eigenvalue $\lambda_i$ of matrix
$\rho^{(D)}$. It follows from this kind of spectrum that
$\rho^{(D)}$ remains invariant under the action of arbitrary
unitary transformation performed in each of the $m_i$-dimensional
eigensubspaces. Therefore $H=U(m_1)\otimes U(m_2)\otimes \cdots
U(m_k)$ is maximum stability subgroup for $\rho^{(D)}$, and the
quotient space $U(n)/H$ is a complex flag manifold
\begin{equation}
{\mathcal F}=\frac{U(n)}{U(m_1)\otimes U(m_2)\otimes \cdots
U(m_k)},\qquad m_1+m_2+\cdots + m_k=N.
\end{equation}
Two special kinds of degeneracy of the spectrum of $\rho^{(D)}$
are as follows: i) Let $\rho^{(D)}$ represents the maximally mixed
state
$\rho_\ast=\textmd{diag}\{\frac{1}{N},\frac{1}{N},\cdots,\frac{1}{N}\}$.
In this case the stability subgroup $H$ is $U(N)$, and the orbit
of point $\rho_\ast$ is only one point, i.e. $\rho=\rho_\ast$. ii)
On the other hand if the spectrum of $\rho^{(D)}$ is
non-degenerate, then the stability subgroup is $n$-dimensional
torus $T^N=U(1)^{\otimes N}$, and the orbit of the point
$\rho^{(D)}$ is
\begin{equation}
\rho={\bf \Omega} \rho^{(D)} {\bf \Omega}^\dag, \qquad {\bf
\Omega}\in U(N)/T^N .
\end{equation}
Since the maximal torus  $T^N$ is itself a subgroup of all maximum
stability subgroups, therefore the orbit of points $\rho^{(D)}\in
{\mathcal D}_N$ under the action of quotient $U(N)/T^N$ generates
all points of the space ${\mathcal M}_N$. The diagonal matrix
$\rho^{(D)}$ is defined up to a permutation of its entries and,
one can divide the simplex ${\mathcal S}_{N-1}$ into $n!$
identical simplexes and take any of them. Each part identifies
points of ${\mathcal S}_{N-1}$ which have the same coordinates,
but with different ordering, and can be considered as the
homomorphic image of simplex ${\mathcal S}_{N-1}$ relative to the
discrete permutation group $P_N$, i.e. ${\mathcal S}_{N-1}/P_N$.
Therefore the points of ${\mathcal M}_N$ can be characterized as
the orbit of diagonal matrices $D\in {\mathcal S}_{N-1}/P_N$ under
the action of quotient ${\bf \Omega}\in U(N)/T^N$.

Further insight into the space of density matrices can be obtained
by writing the quotient ${\bf \Omega}\in U(N)/T^N$ as the product
of $N-1$ components as (\cite{gilmore}, page 401)
\begin{equation}
{\bf \Omega}=\Omega^{(N;N)} \Omega^{(N-1;N)}\cdots \Omega^{(2;N)}.
\end{equation}
where
\begin{equation}
\Omega^{(m;N)}\in \frac{U(m)\otimes T^{N-m}}{U(m-1)\otimes
T^{N-m+1}}, \qquad m=2,\cdots,N.
\end{equation}
A typical coset representative $\Omega^{(m;N)}$ can be written as
\begin{equation}
\Omega^{(m;N)}= \left(
\begin{array}{c|c}
SU(m)/U(m-1) & O \\  \hline  O^T & I_{N-m}
\end{array}
\right),
\end{equation}
where $O$, $O^T$ and $I_{N-m}$ represent, respectively, the
$m\times (N-m)$ zero matrix, its transpose and the $(N-m)\times
(N-m)$ identity matrix. The $2(m-1)$-dimensional coset space
$SU(m)/U(m-1)$ has the following $m\times m$ matrix representation
(\cite{gilmore}, page 351)
\begin{equation}\label{cosetB}
SU(m)/U(m-1)= \left(
\begin{array}{c|c}
\cos{\sqrt{B^{(m)} [B^{(m)}]^\dag}} &
B^{(m)}\frac{\sin{\sqrt{[B^{(m)}]^{\dag}
B^{(m)}}}}{\sqrt{[B^{(m)}]^{\dag}B^{(m)}}} \\  \hline
-\frac{\sin{\sqrt{[B^{(m)}]^{\dag}
B^{(m)}}}}{\sqrt{[B^{(m)}]^{\dag}B^{(m)}}}[B^{(m)}]^\dag
&\cos{\sqrt{[B^{(m)}]^{\dag} B^{(m)}}} \\
\end{array}
\right),
\end{equation}
where $B^{(m)}$ represents an $(m-1)\times 1$ complex matrix and
$[B^{(m)}]^\dag$ is its adjoint.

Now by parameterizing  the column matrix $B^{(m)}$ as
$(\gamma^{(m)}_1{\rm e}^{i\xi^{(m)}_1},\gamma^{(m)}_2{\rm
e}^{i\xi^{(m)}_2},\cdots,\gamma^{(m)}_{m-1}{\rm
e}^{i\xi^{(m)}_{m-1}})^T$ for $m=2,3,\cdots,N$, where
$\gamma^{(m)}_{i}$ and $\xi^{(m)}_{i}$ are real
numbers\footnote{The correspondence between general notation of
this paper and that in Ref. \cite{akhtar2}, where the $N=3$ case
is explicitly computed, is as $\gamma^{(2)}_1=\alpha$,
$\xi^{(2)}_1=\phi$, $\gamma^{(3)}_{1,2}=\beta_{1,2}$ and
$\xi^{(3)}_{1,2}=\psi_{1,2}$.}, the component $\Omega^{(m;N)}$ can
be factorized as
\begin{equation}\label{OXRXd}
\Omega^{(m;N)}=X^{(m;N)}R^{(m;N)}{X^{(m;N)}}^\dag \qquad {\rm for}
\quad m=2,3,\cdots,N,
\end{equation}
where $X^{(m;N)}$ is a  diagonal  $N\times N$ phase matrix with
$X^{(m;N)}_{kl}=\delta_{kl}{\rm exp}\{i\xi^{(m)}_k\}$ and
$\xi^{(m)}_i=0$ for $i\ge m$, and $R^{(m;N)}$ is an $N\times N$
orthogonal matrix with the following nonzero elements
$$
R^{(m;N)}_{ij}=\delta_{ij} + {\hat \gamma}^{(m)}_i{\hat
\gamma}^{(m)}_j (\cos{\gamma^{(m)}}-1) \qquad {\rm for} \quad
1\le i,j\le m-1
$$\vspace{-2mm}
$$ \hspace{-9mm}
R^{(m;N)}_{im}=-R^{(m;N)}_{mi}={\hat \gamma}^{(m)}_i
\sin{\gamma^{(m)}}  \qquad{\rm for} \quad 1\le i\le m-1
$$\vspace{-2mm}
$$\hspace{-71mm}
R^{(m;N)}_{mm}=\cos{\gamma^{(m)}}
$$\vspace{-2mm}
\begin{equation}\label{RmN}\hspace{-44mm}
R^{(m;N)}_{ii}=1   \qquad{\rm for} \quad m+1\le i\le N
\end{equation}
where we have defined ${\hat
\gamma}^{(m)}_i=\gamma^{(m)}_i/\gamma^{(m)}$ and
$\gamma^{(m)}=\sqrt{\sum_{i=1}^{m-1}(\gamma^{(m)}_i)^2}$. As we
will see later the important ingredient of our approach in
computing  the Bures metric is the possibility of writing the
factorization (\ref{OXRXd}).

\section{Bures metric}
In this section we shall attempt to develop a method of computing
the Bures metric of an arbitrary density matrix of an $N$-level
quantum system.  We will use the canonical coset parametrization
of the density matrices introduced in the last section.

Let $\rho$ be a generic density matrix of an $N$-level quantum
system, with eigenvalues $\lambda_j$ and corresponding
eigenvectors $|\lambda_j\rangle$, $(j=1,2,\cdots,N)$. In the light
of $\rho={\bf \Omega} \rho^{(D)} {\bf \Omega}^\dag $, with
$\rho^{(D)}=\textmd{diag}\{\lambda_1,\lambda_2,\cdots,\lambda_N\}$
as the diagonal matrix of $\rho$ eigenvalues, the $\rho$
eigenvectors can be written in terms of $\rho^{(D)}$ eigenvectors
as $|\lambda_i\rangle={\bf \Omega}|i\rangle$. Therefore invoking
the H\"{u}bner formula (\ref{hubner}), we can write the
infinitesimal Bures distance between $\rho$ and
$\rho+\textmd{d}\rho$ as
\begin{equation}\label{hubner2}
\textmd{d}_B(\rho,\rho+\d\rho)^2=\frac{1}{2}\sum_{i,j=1}^{N}\frac{|\left(\langle
i|{\bf \Omega}^\dag\right)\;\textmd{d} \left({\bf
\Omega}\rho^{(D)}{\bf \Omega}^\dag\right)\left({\bf
\Omega}|j\rangle\right)|^2}{\lambda_i+\lambda_j},
\end{equation}
which takes the form (remember that ${\bf \Omega}$ is unitary and
therefore ${\rm d} {\bf \Omega}^\dag=-{\bf \Omega}^\dag\;{\rm
d}{\bf \Omega}\;{\bf \Omega}^\dag$)
\begin{equation}
\d_B(\rho,\rho+{\rm
d}\rho)^2=\frac{1}{2}\sum_{i,j=1}^{N}\frac{|\langle i|{\rm
d}\rho^{(D)}|j\rangle+ \langle i| [{\bf \Omega}^\dag \d {\bf
\Omega},\rho^{(D)}]|j\rangle|^2}{\lambda_i+\lambda_j},
\end{equation}
By using the equations $\rho^{(D)}|i\rangle=\lambda_i|i\rangle$
and  $\langle i|j\rangle=\delta_{ij}$ we get
\begin{equation}\label{dB=dD+dC}
\d_B(\rho,\rho+\d\rho)^2=\sum_{i=1}^{N}\frac{\left({\rm
d}\lambda_i\right)^2}{4\lambda_i}+\sum_{i<j}^{N}\Lambda_{ij}|({\bf
\Omega}^\dag{\rm d} {\bf \Omega})_{ij}|^2,
\end{equation}
where we have defined $\Lambda_{ij}$ as
\begin{equation}
\Lambda_{ij}=\frac{\left(\lambda_i-\lambda_j\right)^2}
{\lambda_i+\lambda_j}.
\end{equation}
Equation (\ref{dB=dD+dC}) shows that the infinitesimal Bures
distance is divided into two infinitesimal distances corresponding
to the eigenvalues coordinates and coset coordinates. Therefore
the Bures metric matrix  becomes  block diagonal as
\begin{equation} g=\left(
\begin{array}{c|c}
 g^{(D)} & 0 \\
 \hline
 0 & g^{(C)}
\end{array}  \right),
\end{equation}
where $g^{(D)}$ is a part of the Bures metric corresponding to the
diagonal density matrix $\rho^{(D)}$, and $g^{(C)}$ is the
contribution of the coset coordinates in the Bures metric.

In what follows our goal is to calculate the matrix elements of
$g^{(D)}$ and $g^{(C)}$. In order to calculate $g^{(D)}$, we first
note that the $N$ eigenvalues can be parameterized explicitly in
terms of $N-1$ independent parameters $\theta_k$
$(k=1,2,\cdots,N-1)$ as
\begin{equation}
\left\{\begin{array}{ll}
\lambda_{k}=\sin^2{\theta_1}\sin^2{\theta_2}\cdots\sin^2{\theta_{k-1}}\cos^2{\theta_k}
& {\rm for} \quad k=1,2,\cdots,N-1 \\ \\
\lambda_{N}=\sin^2{\theta_1}\sin^2{\theta_2}\cdots\sin^2{\theta_{N-1}}.
&
\end{array}\right.
\end{equation}
By using the above coordinates for the eigenvalues of $\rho$, we
can write the infinitesimal Bures distance over the eigenvalues
coordinates as
\begin{equation}
\sum_{i=1}^{N}\frac{\left({\rm
d}\lambda_i\right)^2}{4\lambda_i}=\sum_{k,l=1}^{N-1}g_{kl}^{(D)}{\rm
d}\theta_k{\rm d}\theta_l
\end{equation}
where it can be easily seen that the metric $g_{kl}^{(D)}$ is
diagonal with elements
$$\hspace{-82mm}
g^{(D)}_{\theta_1\theta_1}=1
$$
\begin{equation}
g^{(D)}_{\theta_k\theta_k}=\sin^2{\theta_1}\sin^2{\theta_2}\cdots\sin^2{\theta_{k-1}},
\qquad {\rm for} \quad k=2,3,\cdots,N-1.
\end{equation}
It is worth noting that the metric $g^{(D)}$ is independent of
$\theta_{N-1}$.

 Now in order to calculate $g^{(C)}$, we define
$\chi^{(m)}_{\alpha_m}$ $(\alpha_m=1,2,\cdots,2(m-1))$ as the
$2(m-1)$ real parameters of the coset component $\Omega^{(m;N)}$
$(m=2,\cdots,N)$ such that
\begin{equation}
\chi^{(m)}_{\alpha_m}=\left\{\begin{array}{ll}
\gamma^{(m)}_{\alpha_m} & {\rm for} \quad \alpha_m=1,\cdots,m-1,
\\ \\ \xi^{(m)}_{\alpha_m} & {\rm for} \quad
\alpha_m=m,\cdots,2(m-1),
\end{array}\right.
\end{equation}
Then the infinitesimal Bures distance over the coset coordinates
can be written as
\begin{equation}
\begin{array}{rl} \vspace{3mm}
\sum_{i<j}^{N}\Lambda_{ij}|({\bf \Omega}^\dag{\rm d} {\bf
\Omega})_{ij}|^2= &
\sum_{m=2}^{N}\sum_{m^\prime=2}^{N}\sum_{\alpha_m=1}^{2(m-1)}
\sum_{\beta_{m^\prime}
=1}^{2(m^\prime-1)}g_{\alpha_m,\beta_{m^\prime}}^{(m,m^\prime;N)}{\rm
d}\chi_{\alpha_m}^{(m)}\d\chi_{\beta_{m^\prime}}^{(m^\prime)}\\
\vspace{3mm}
 = &
\sum_{m=2}^{N}\sum_{\alpha_m,\beta_m =1}^{2(m-1)}
g_{\alpha_m,\beta_m}^{(m,m;N)}{\rm
d}\chi_{\alpha_m}^{(m)}\d\chi_{\beta_m}^{(m)} \\
+ & 2\sum_{m<m^\prime}^{N}
\sum_{\alpha_m=1}^{2(m-1)}\sum_{\beta_{m^\prime}
=1}^{2(m^\prime-1)}
g_{\alpha_m,\beta_{m^\prime}}^{(m,{m^\prime};N)}{\rm
d}\chi_{\alpha_m}^{(m)}\d\chi_{\beta_{m^\prime}}^{(m^\prime)}.
\end{array}
\end{equation}
Now invoking the decomposition
\begin{equation}
{\bf \Omega}=\Omega^{(N;N)} \Omega^{(N-1;N)}\cdots \Omega^{(2;N)},
\end{equation}
we can write
\begin{equation}\label{OdagmdOmij}
|({\bf \Omega}^\dag\d {\bf \Omega})_{ij}|^2=
\sum_{m=2}^{N}\left|\left(K^{(m;N)}\right)_{ij}\right|^2  +
2\sum_{m<m^\prime}^{N}{\rm
Re}\left\{\left(K^{(m;N)}\right)_{ij}\left(K^{(m^\prime;N)}\right)_{ij}^{\ast}\right\},
\end{equation}
where we have defined
\begin{equation}\label{KmN}
K^{(m;N)}={W^{(m;N)}}^\dag \left({\Omega^{(m;N)}}^\dag
\d\Omega^{(m;N)}\right)W^{(m;N)},
\end{equation}
with
\begin{equation}
W^{(m;N)}=\Omega^{(m-1;N)}\cdots\Omega^{(3;N)}\Omega^{(2;N)}
\end{equation}
and $W^{(2;N)}=1$. The first term of the r.h.s. of  equation
(\ref{OdagmdOmij}) shows sum over all pure contribution of each
coset component in the Bures distance, and can be identified with
\begin{equation}\label{gchi11}
\sum_{i<j}^{n}\Lambda_{ij}\left|\left(K^{(m;N)}\right)_{ij}\right|^2=\sum_{\alpha_m,\beta_m
=1}^{2(m-1)}
g_{\alpha_m,\beta_m}^{(m,m;N)}\d\chi_{\alpha_m}^{(m)}{\rm
d}\chi_{\beta_m}^{(m)}
\end{equation}
The second term, however, shows the sum over  mixed contribution
of all pairs of coset components in the Bures distance and  can be
used to write
\begin{equation}\label{gchi12}
\sum_{i<j}^{n}\Lambda_{ij}  {\rm
Re}\left\{\left(K^{(m;N)}\right)_{ij}\left(K^{(m^\prime;N)}\right)_{ij}^{\ast}\right\}
=\sum_{\alpha_m,=1}^{2(m-1)}\sum_{\beta_{m^\prime}
=1}^{2(m^\prime-1)}
g_{\alpha_m,\beta_{m^\prime}}^{(m,m^\prime;N)}{\rm
d}\chi_{\alpha_m}^{(m)}\d\chi_{\beta_{m^\prime}}^{(m^\prime)}.
\end{equation} Therefore  $g^{(C)}$ is defined by the
following symmetric matrix
\begin{equation}
g^{(C)}=\left(
 \begin{array}{c|c|c|c}
 g^{(2,2;N)} & g^{(2,3;N)}
&  \cdots & g^{(2,N;N)}
  \\  \hline
  &g^{(3,3;N)}  &
\cdots & g^{(3,N;N)}
 \\
\hline & &
  \ddots & \vdots \\
\hline & & &  g^{(N,N;N)}
\end{array}\right).
\end{equation}
Now the object is to calculate the matrix elements of $g^{(C)}$,
which can be achieved if we can find an expression for
$\left(K^{(m;N)}\right)_{ij}$ of equation (\ref{gchi11}),
(\ref{gchi12}) in terms of the coset coordinates
$\chi^{(m)}_{\alpha_m}$(or equivalently in terms of
$\gamma^{(m)}_{\alpha_m}$ and $\xi^{(m)}_{\alpha_m}$). Therefore
we have to expand ${\Omega^{(m;N)}}^\dag \d\Omega^{(m;N)}$ of
equation (\ref{KmN}) in terms of $\d\gamma^{(m)}$ and
$\d\xi^{(m)}$. This can be achieved by using the factorization
$\Omega^{(m;N)}=X^{(m;N)}R^{(m;N)}{X^{(m;N)}}^\dag$ where we get
\begin{equation}\label{OdagmdOm}
\begin{array}{rl}
{\Omega^{(m;N)}}^\dag {\rm d}\Omega^{(m;N)}= &
X^{(m;N)}{R^{(m;N)}}^T {X^{(m;N)}}^{\dag}({\rm
d}X^{(m;N)})R^{(m;N)}{X^{(m;N)}}^\dag \\
- & ({\rm d}X^{(m;N)}){X^{(m;N)}}^\dag \\
+ & X^{(m;N)}{R^{(m;N)}}^T({\rm d}R^{(m;N)}){X^{(m;N)}}^\dag.
\end{array}
\end{equation}
By knowing that $X^{(m;N)}$ is a unitary diagonal matrix and
$R^{(m;N)}$ is an orthogonal matrix, the first two terms can be
calculated easily by using the relation
\begin{equation}\label{dX}
{\rm d}X^{(m;N)}_{ij}=i{\rm e}^{i\xi^{(m)}_i}\delta_{ij}{\rm
d}\xi^{(m)}_i.
\end{equation}
 On the other hand to calculate
the third term, the equation (\ref{RmN}) may be used to show that
\begin{equation}\label{dR}
{\rm d}R^{(m;N)}_{ij}=\sum_{k=1}^{m-1}\Gamma^{(m,N)}_{ij;k}{\rm
d}\gamma^{(m)}_k,
\end{equation}
where for a fixed $k$ $(=1,2,\cdots,m-1)$, the
$\Gamma^{(m,N)}_{ij;k}$ denote the matrix elements of an $N\times
N$ matrix with nonzero elements
$$
\Gamma^{(m;N)}_{ij;k}= \left({\hat
\gamma}^{(m)}_i\delta_{jk}+{\hat \gamma}^{(m)}_j
\delta_{ik}-2{\hat \gamma}^{(m)}_i{\hat \gamma}^{(m)}_j{\hat
\gamma}^{(m)}_k\right)(\cos{\gamma^{(m)}}-1)/\gamma^{(m)}
$$
$$\hspace{-6mm}
-{\hat \gamma}^{(m)}_i{\hat \gamma}^{(m)}_j{\hat
\gamma}^{(m)}_k\sin{\gamma^{(m)}} \qquad {\rm for} \quad 1\le
i,j\le m-1
$$
\vspace{2mm}
$$\hspace{-24mm}
\Gamma^{(m;N)}_{im;k}=-\Gamma^{(m;N)}_{mi;k}=
\left(\delta_{ik}-{\hat \gamma}^{(m)}_i{\hat
\gamma}^{(m)}_k\right)\sin{\gamma^{(m)}}/\gamma^{(m)}
$$
$$\hspace{-16mm}
+{\hat \gamma}^{(m)}_i{\hat \gamma}^{(m)}_k\cos{\gamma^{(m)}}
\qquad {\rm for} \quad 1\le i\le m-1
$$
\vspace{2mm}
\begin{equation}\hspace{-68mm}
\Gamma^{(m;N)}_{mm;k}=-{\hat \gamma}^{(m)}_k \sin{\gamma^{(m)}}
\end{equation}
Using equations (\ref{dX}) and  (\ref{dR}) in equation
(\ref{OdagmdOm}), we get
\begin{equation}
\begin{array}{rl}
\left({\Omega^{(m;N)}}^\dag \d\Omega^{(m;N)}\right)_{ij}= & {\rm
exp}\{i(\xi^{(m,N)}_i-\xi^{(m,N)}_j)\} \\
\times &
\sum_{k=1}^{m-1}\left(\left(E^{(m;N)}_{ij}\right)_{\gamma_k}{\rm
d}\gamma^{(m)}_k+i \left(E^{(m;N)}_{ij}\right)_{\xi_k}{\rm
d}\xi^{(m)}_k\right)
\end{array}
\end{equation}
where $\left(E^{(m;N)}_{ij}\right)_{\gamma_k}$ and
$\left(E^{(m;N)}_{ij}\right)_{\xi_k}$ have been defined by
\begin{equation}
\left(E^{(m;N)}_{ij}\right)_{\gamma_k}=\sum_{l=1}^{m}R^{(m;N)}_{li}\Gamma^{(m;N)}_{lj;k}
\end{equation}
\begin{equation}
\left(E^{(m;N)}_{ij}\right)_{\xi_k}=R^{(m;N)}_{ki}R^{(m;N)}_{kj}+\delta_{ij}\delta_{jk}
\end{equation}
Therefore for $K^{(m;N)}$ of equation (\ref{KmN}), we find the
following matrix elements
\begin{equation}
K^{(m;N)}_{ij}=\sum_{r=1}^{m-1}\left\{\left(K^{(m;N)}_{ij}\right)_{\gamma_r}{\rm
d}\gamma^{(m)}_r+\left(K^{(m;N)}_{ij}\right)_{\xi_r}{\rm
d}\xi^{(m)}_r\right\},
\end{equation}
where we have defined
\begin{equation}
\left(K^{(m;N)}_{ij}\right)_{\gamma_r}=\sum_{k,l=1}^{m}\left(W^{(m;N)}_{ik}\right)^\ast
\left(W^{(m;N)}_{lj}\right)\left(E^{(m;N)}_{kl}\right)_{\gamma_r}.
\end{equation}
and
\begin{equation}
\left(K^{(m;N)}_{ij}\right)_{\xi_r}=\sum_{k,l=1}^{m}\left(W^{(m;N)}_{ik}\right)^\ast
\left(W^{(m;N)}_{lj}\right)\left(E^{(m;N)}_{kl}\right)_{\xi_r},
\end{equation}
Finally, putting everything together, we find the following
formula for the matrix elements of the metric $g^{(C)}$
\begin{equation}\label{ggg}
g_{\gamma_r^{(m)}\gamma_s^{(m^\prime)}}^{(m,m^\prime;N)}=\sum_{i<j}^{N}\Lambda_{ij}
\;{\rm Re} \left\{\left(K^{(m;N)}_{ij}\right)_{\gamma_r^{(m)}}
\left(K^{(m^\prime;N)}_{ij}\right)_{\gamma_s^{(m^\prime)}}^\ast
\right\},
\end{equation}
\begin{equation}\label{gxx}
g_{\xi_r^{(m)}\xi_s^{(m^\prime)}}^{(m,m^\prime;N)}=\sum_{i<j}^{N}\Lambda_{ij}
\;{\rm Re} \left\{\left(K^{(m;N)}_{ij}\right)_{\xi_r^{(m)}}
\left(K^{(m^\prime;N)}_{ij}\right)_{\xi_s^{(m^\prime)}}^\ast
\right\},
\end{equation}
\begin{equation}\label{ggx}
g_{\gamma_r^{(m)}\xi_s^{(m^\prime)}}^{(m,m^\prime;N)}=\sum_{i<j}^{N}\Lambda_{ij}
\;{\rm Re} \left\{\left(K^{(m;N)}_{ij}\right)_{\gamma_r^{(m)}}
\left(K^{(m^\prime;N)}_{ij}\right)_{\xi_s^{(m^\prime)}}^\ast
\right\},
\end{equation}
\begin{equation}\label{gxg}
g_{\xi_r^{(m)}\gamma_s^{(m^\prime)}}^{(m,m^\prime;N)}=\sum_{i<j}^{N}\Lambda_{ij}
\;{\rm Re} \left\{\left(K^{(m;N)}_{ij}\right)_{\xi_r^{(m)}}
\left(K^{(m^\prime;N)}_{ij}\right)_{\gamma_s^{(m^\prime)}}^\ast
\right\},
\end{equation}
for $r=1,2,\cdots,m-1$, $s=1,2,\cdots,m^\prime-1$ and
$m,m^\prime=1,2,\cdots,N$. Since the eigenvalues coordinates are
just included  in the $\Lambda_{ij}$ terms of  equations
(\ref{ggg})-(\ref{gxg}), therefore it is clear that all matrix
elements of the metric $g^{(C)}$ are simply sum of the products of
two independent functions,  the eigenvalues coordinates and the
coset coordinates. It should be noted that for real density
matrices, i.e. $\xi^{(m)}_i=0$ for $i=1,2,\cdots,m-1$ and
$m=1,2,\cdots,N$,  we have $\Omega^{(m;N)}=R^{(m;N)}$. In this
case all quantities are real and equation (\ref{ggg}) gives all
matrix elements of the $N(N-1)/2$-dimensional metric $g^{(C)}$.

\section{Conclusion}
We provide a method for explicit  computation of the Bures metric
in $N$-level quantum systems, based on the coset parametrization
of density matrices. We show that in the canonical coset
parametrization, the Bures metric matrix is divided into two
blocks, an $(N-1)$-dimensional  diagonal  matrix corresponding to
the eigenvalues coordinates, and an $N(N-1)$-dimensional matrix
corresponding to the coset coordinates. It also provides a
factorization of the Bures measure on the space of density
matrices as the product of the measure on the space of eigenvalues
and the truncated Haar measure on the space of unitary matrices.

\end{document}